\documentclass[aps,pra,showpacs,superscriptaddress,preprint]{revtex4}
\usepackage{graphicx}
\catcode`ð=\active
\defð{\u{g}}
\catcode`Ð=\active
\defÐ{\u{G}}
\catcode`Ý=\active
\defÝ{\. I}
\catcode`ö=\active
\defö{\"{o}}
\catcode`Ö=\active
\defÖ{\"O}
\catcode`ü=\active
\defü{\"{u}}
\catcode`Ü=\active
\defÜ{\"{U}}
\catcode`Þ=\active
\defÞ{\c{S}}
\catcode`þ=\active
\defþ{\c{s}}
\catcode`ý=\active
\defý{{\i}}
\catcode`ç=\active
\defç{\d{c}}
\catcode`Ç=\active
\defÇ{\d{C}}

\begin{document}
\title{Approximate Pseudospin and Spin Solutions of the Dirac
Equation for a Class of Exponential Potentials}
\author{\small Altuð Arda}
\email[E-mail: ]{arda@hacettepe.edu.tr}\affiliation{Department of
Physics Education, Hacettepe University, 06800, Ankara,Turkey}
\author{\small Ramazan Sever}
\email[E-mail: ]{sever@metu.edu.tr}\affiliation{Department of
Physics, Middle East Technical  University, 06800, Ankara,Turkey}
\author{\small Cevdet Tezcan}
\email[E-mail: ]{ctezcan@baskent.edu.tr}\affiliation{Faculty of
Engineering, Baþkent University, Baglýca Campus, Ankara,Turkey}
\begin{abstract}
Dirac equation is solved for some exponential potentials,
hypergeometric-type potential, generalized Morse potential and
Poschl-Teller potential with any spin-orbit quantum number $\kappa$
in the case of spin and pseudospin symmetry, respectively. We have
approximated for non s-waves the centrifugal term by an exponential
form. The energy eigenvalue equations, and the corresponding wave
functions are obtained by using the
generalization of the Nikiforov-Uvarov method.\\
Keywords: Pseudospin symmetry, Spin symmetry, Dirac Equation,
Hypergeometric Potential, generalized Morse Potential, Poschl-Teller
Potential, Nikiforov-Uvarov Method
\end{abstract}

\pacs{03.65.Fd, 03.65.Ge, 12.39.Fd}
\maketitle
\newpage
\section{Introduction}
The solutions of the Dirac equation having the pseudospin, and
spin symmetry have been strongly studied in the last years. The
concept of the pseudospin symmetry [1, 2, 3] is an important theme
in nuclear theory because of its features related to construct an
effective shell-model coupling scheme [4], to study the structures
of the deformed nuclei [5, 6]. The pseudopin symmetry occurs in
nuclei when the magnitude of the scalar and vector potentials are
nearly equal, but opposite sign, i.e., $V_{v}(r) \sim
-V_{s}(r)$\,,where the scalar potential is negative (attractive)
and the vector potential is positive (repulsive) in the
relativistic region. Further, the spin symmetry appears when the
magnitude of the scalar and vector potentials are nearly equal,
i.e., $V_{v}(r) \sim V_{s}(r)$ [7-12]. The Dirac equation under
the exact pseudospin and/or spin symmetry has been studied by with
different type of potentials such as the Hulth\'{e}n potential
[13], the Morse potential [14], the Woods-Saxon potential [15,
16], and harmonic oscillator [17-21].

In the present work, we deal with the solutions of the Dirac
equation if the exact pseudospin, and spin symmetry occur in the
theory under the effect of a class of potentials, which have
exponential form, i.e. the hypergeometric-type potential [22], the
generalized Morse potential and the Pöschl-Teller potential [23].

In order to obtain the energy eigenvalue equation and the
corresponding wave functions, we apply a new approximation scheme
[24], which is the parametric generalization of the
Nikiforov-Uvarov (NU) method [25], by using an approximation to
the centrifugal-like term. So, we obtain the energy spectra of the
above potentials for the spin-orbit quantum number $\kappa=0$, or
for any  $\kappa$-value for the case of pseudospin and spin
symmetry, respectively.

The organization of the present work is as follows. Firstly, we
give briefly the equations for the Dirac spinors including the
centrifugal term, and the spin-orbit quantum number $\kappa$. Than
we present the basics of the parametric generalization of the NU
method. We find the bound states and the corresponding wave
functions of the Dirac equation with the above potentials in the
case of pseudospin and spin symmetry, respectively. Finally, we
give our conclusions.

\section{Dirac Equation}
The Dirac equation for a particle with rest mass $m$ in the
absence of the scalar and vector potential is written
$(\hbar=c=1)$ [15]

\begin{eqnarray}
\Big\{\vec{\alpha}.\vec{p}+\beta[m+V_{s}(r)]\Big\}\Psi(r)=
(E-V_{v}(r))\Psi(r)\,,
\end{eqnarray}
where $\vec{p}$ is the momentum operator, $\vec{\alpha}$, and
$\beta$ are $4 \times 4$ Dirac matrices, i.e. $\vec{\alpha}=\left(
\begin{array}{cc}
0 & \vec{\sigma}_{i}\\
\vec{\sigma}_{i} & 0 \\
\end{array}\right)\,\,,$ $\beta=\left(
\begin{array}{cc}
I & 0\\
0 & -I \\
\end{array}\right)\,,$
$I$ is the $2 \times 2$ unit matrix, and $\vec{\sigma}_{i}(i=1, 2,
3)$ are the Pauli matrices. The operator $\hat{K}$ is the
spin-orbit matrix operator and written in terms of the orbital
angular momentum operator $\hat{L}$ as
$\hat{K}=-\beta(\hat{\sigma}.\hat{L}+1)$\,,which commute with the
Dirac Hamiltonian. The Dirac spinors can be labelled by the
quantum number set $(n,\kappa)$\,,where $\kappa$ is the eigenvalue
of the spin-orbit operator, and written as

\begin{eqnarray}
\Psi_{n\kappa}(r)=\left(
\begin{array}{c}
f_{n\kappa}\\
g_{n\kappa} \\
\end{array}\right)\,,
\end{eqnarray}
where $f_{n\kappa}=[F_{n\kappa}(r)/r]Y^{\ell}_{jm}(\theta,\phi)$
is the upper, and
$g_{n\kappa}=[iG_{n\kappa}(r)/r]Y^{\tilde{\ell}}_{jm}(\theta,\phi)$
is the lower component, and $Y^{\ell}_{jm}(\theta,\phi)$\,, and
$Y^{\tilde{\ell}}_{jm}(\theta,\phi)$ are the spherical harmonics,
respectively. The total angular momentum, the orbital angular
momentum, and pseudo-orbital angular momentum can be written in
terms of the spin-orbit quantum number $\kappa=\pm 1, \pm
2,\ldots$\,, such as $j=|\kappa|-1/2$\,,
$\ell=|\kappa+1/2|-1/2$\,, and $\tilde{\ell}=|\kappa-1/2|-1/2$\,,
respectively.

Substituting Eq. (2) into Eq. (1), and eliminating
$F_{n\kappa}(r)$, we obtain two uncoupled differential equations
for the lower, and upper components of the Dirac equation

\begin{eqnarray}
\Big\{\frac{d^2}{dr^2}-\frac{\kappa(\kappa-1)}{r^2}-\frac{1}{M_{\Sigma}(r)}
\frac{d\Sigma(r)}{dr}\Big(\frac{d}{dr}-\frac{\kappa}{r}\Big)\Big\}G_{n\kappa}(r)
=M_{\Delta}(r)M_{\Sigma}(r)G_{n\kappa}(r)\,,
\end{eqnarray}
and

\begin{eqnarray}
\Big\{\frac{d^2}{dr^2}-\frac{\kappa(\kappa+1)}{r^2}+\frac{1}{M_{\Delta}(r)}
\frac{d\Delta(r)}{dr}\Big(\frac{d}{dr}+\frac{\kappa}{r}\Big)\Big\}F_{n\kappa}(r)
=M_{\Delta}(r)M_{\Sigma}(r)F_{n\kappa}(r)\,.
\end{eqnarray}
where $M_{\Delta}(r)=m+E-\Delta(r)$\,,
$\Delta(r)=V_{v}(r)-V_{s}(r)$\,, $M_{\Sigma}(r)=m-E+\Sigma(r)$\,,
and $\Sigma(r)=V_{v}(r)+V_{s}(r)$\,.

\section{Parametric Generalization of the Method}

Let us give briefly the parametric generalization of the NU
method. By using an appropriate coordinate transformation, the
Schr\"{o}dinger equation can be transformed into the following
form

\begin{eqnarray}
\sigma^2(z)\frac{d^2\Psi(z)}{dz^2}+\sigma(z)\tilde{\tau}(z)\frac{d\Psi(z)}{dz}+\tilde{\sigma}(z)
\Psi(z)=0\,,
\end{eqnarray}
where $\sigma(z)$\,, and $\tilde{\sigma}(z)$ are polynomials, at
most, second degree, and $\tilde{\tau}(z)$ is a first-degree
polynomial. By writing the general solution as
$\Psi(z)=\psi(z)\varphi(z)$, we obtain a hypergeometric type
equation [24]

\begin{eqnarray}
\frac{d^2\varphi(z)}{dz^2}+\frac{\tau(z)}{\sigma(z)}\frac{d\varphi(z)}{dz}+\frac{\lambda}{\sigma(z)}\,
\varphi(z)=0\,,
\end{eqnarray}
where $\psi(z)$ and $\varphi_{n}(z)$ are defined as [24]

\begin{eqnarray}
\frac{1}{\psi(z)}\frac{d\psi(z)}{dz}=\frac{\pi(z)}{\sigma(z)}\,,
\end{eqnarray}

\begin{eqnarray}
\varphi_{n}(z)=\frac{a_{n}}{\rho(z)}\frac{d^n}{dz^n}[\sigma^{n}(z)\rho(z)]\,,
\end{eqnarray}
where $a_{n}$ is a normalization constant, and $\rho(z)$ is the
weight function satisfying the following equation [24]

\begin{eqnarray}
\frac{d\sigma(z)}{dz}+\frac{\sigma(z)}{\rho(z)}\frac{d\rho(z)}{dz}=\tau(z)\,.
\end{eqnarray}
The function $\pi(z)$\,, and the parameter $\lambda$ in the above
equation are defined as

\begin{eqnarray}
\pi(z)&=&\,\frac{1}{2}\,[\sigma'(z)-\tilde{\tau}(z)]\pm\Bigg[\frac{1}{4}[\sigma'(z)-\tilde{\tau}(z)]^2
-\tilde{\sigma}(z)+k\sigma(z)\Bigg]^{1/2}\,,\\
\lambda&=&k+\pi'(z)\,.
\end{eqnarray}

In the NU method, the square root in Eq. (10) must be the square
of the polynomial, so the parameter $k$ can be determined. Thus, a
new eigenvalue equation becomes

\begin{eqnarray}
\lambda=\lambda_{n}=-n\tau'(z)-\frac{1}{2}(n^2-n)\sigma''(z)\,.
\end{eqnarray}
where prime denotes the derivative and the derivative of the
function $\tau(z)=\tilde{\tau}(z)+2\pi(z)$ should be negative.

Now, in order to clarify the parametric generalization of the NU
method [25], let us take the following Schr\"{o}dinger-like
equation written for any potential

\begin{eqnarray}
z^2(1-\alpha_{3}z)^2\frac{d^2\Psi(z)}{dz^2}+z(1-\alpha_{3}z)(\alpha_{1}-\alpha_{2}z)
\frac{\Psi(z)}{dz}+[-\xi_{1}z^2+\xi_{2}z-\xi_{3}]\Psi(z)=0\,.
\end{eqnarray}

Comparing Eq. (13) with Eq. (5), we obtain

\begin{eqnarray}
\tilde{\tau}(z)=\alpha_{1}-\alpha_{2}z\,\,;\,\sigma(z)=z(1-\alpha_{3}z)\,\,;\,\tilde{\sigma}(z)=
-\xi_{1}z^2+\xi_{2}z-\xi_{3}\,.
\end{eqnarray}

Substituting these into Eq. (10), we obtain

\begin{eqnarray}
\pi(z)=\alpha_{4}+\alpha_{5}z\pm\Bigg[(\alpha_{6}-k\alpha_{3})z^2+(\alpha_{7}+k)z+\alpha_{8}\Bigg]^{1/2}\,,
\end{eqnarray}
with the following parameters

\begin{eqnarray}
\begin{array}{ll}
\alpha_{4}=\frac{1}{2}\,(1-\alpha_{1})\,, & \alpha_{5}=\frac{1}{2}\,(\alpha_{2}-2\alpha_{3})\,, \\
\alpha_{6}=\alpha_{5}^{2}+\xi_{1}\,, &
\alpha_{7}=2\alpha_{4}\alpha_{5}-\xi_{2}\,, \\
\alpha_{8}=\alpha_{4}^{2}+\xi_{3}\,. & \\
\end{array}
\end{eqnarray}

We obtain the parameter $k$ from the condition that the function
under the square root should be the square of a polynomial

\begin{eqnarray}
k_{1,2}=-(\alpha_{7}+2\alpha_{3}\alpha_{8})\pm2\sqrt{\alpha_{8}\alpha_{9}}\,,
\end{eqnarray}
where
$\alpha_{9}=\alpha_{3}\alpha_{7}+\alpha_{3}^{2}\alpha_{8}+\alpha_{6}$\,.
The function $\pi(z)$ becomes

\begin{eqnarray}
\pi(z)=\alpha_{4}+\alpha_{5}z-\left[(\sqrt{\alpha_{9}}+\alpha_{3}
\sqrt{\alpha_{8}}\,)z-\sqrt{\alpha_{8}}\,\right].
\end{eqnarray}
for the $k$-value
$k=-(\alpha_{7}+2\alpha_{3}\alpha_{8})-2\sqrt{\alpha_{8}\alpha_{9}}$\,.
We also have from $\tau(z)=\tilde{\tau}(z)+2\pi(z)$\,,

\begin{eqnarray}
\tau(z)=\alpha_{1}+2\alpha_{4}-(\alpha_{2}-2\alpha_{5})z-2\left[(\sqrt{\alpha_{9}}
+\alpha_{3}\sqrt{\alpha_{8}}\,)z-\sqrt{\alpha_{8}}\,\right].
\end{eqnarray}

Thus, we impose the following condition to fix the $k$-value

\begin{eqnarray}
\tau^{\prime}(z)&=&-(\alpha_{2}-2\alpha_{5})-2(\sqrt{\alpha_{9}}
+\alpha_{3}\sqrt{\alpha_{8}}\,) \nonumber \\[5mm]
&=&-2\alpha_{3}-2(\sqrt{\alpha_{9}}+\alpha_{3}\sqrt{\alpha_{8}}\,)\quad
<0.
\end{eqnarray}

From Eqs. (11), (18) and (19) and by using
$\tau(z)=\tilde{\tau}(z)+2\pi(z)$\ and equating Eq. (11) with the
condition that $\lambda$ must satisfy given by Eq. (12), we obtain
the energy eigenvalue equation for the potential under the
consideration

\begin{eqnarray}
n[(n-1)\alpha_{3}+\alpha_{2}-2\alpha_{5}]&-&\alpha_{5}+(2n+1)
(\sqrt{\alpha_{9}}+\alpha_{3}\sqrt{\alpha_{8}}\,)\nonumber\\&+&\alpha_{7}+
2\alpha_{3}\alpha_{8}+2\sqrt{\alpha_{8}\alpha_{9}\,}=0\,.
\end{eqnarray}

By using Eq. (9)

\begin{eqnarray}
\rho(z)=z^{\alpha_{10}-1}(1-\alpha_{3}z)^{\frac{\alpha_{11}}{\alpha_{3}}-\alpha_{10}-1}\,,
\end{eqnarray}
and together with Eq. (8), we obtain

\begin{eqnarray}
\varphi_{n}(z)=P_{n}^{(\alpha_{10}-1,\frac{\alpha_{11}}{\alpha_{3}}-\alpha_{10}-1)}(1-2\alpha_{3}z)\,,
\end{eqnarray}
where
$\alpha_{10}=\alpha_{1}+2\alpha_{4}+2\sqrt{\alpha_{8}}\,\,,\,
\alpha_{11}=\alpha_{2}-2\alpha_{5}+2(\sqrt{\alpha_{9}}+\alpha_{3}\sqrt{\alpha_{8}}$,
and $P_{n}^{(\alpha,\beta)}(1-2\alpha_{3}z)$ are Jacobi
polynomials. By using Eq. (7), we obtain

\begin{eqnarray}
\psi(z)=z^{\alpha_{12}}(1-\alpha_{3}z)^{-\alpha_{12}-\frac{\alpha_{13}}{\alpha_{3}}}\,,
\end{eqnarray}
and the total wave function become

\begin{eqnarray}
\Psi(z)=z^{\alpha_{12}}(1-\alpha_{3}z)^{-\alpha_{12}-\frac{\alpha_{13}}{\alpha_{3}}}
P_{n}^{(\alpha_{10}-1,\frac{\alpha_{11}}{\alpha_{3}}-\alpha_{10}-1)}(1-2\alpha_{3}z)\,,
\end{eqnarray}
where
$\alpha_{12}=\alpha_{4}+\sqrt{\alpha_{8}}\,\,,\,\alpha_{13}=\alpha_{5}-(\sqrt{\alpha_{9}}+\alpha_{3}\sqrt{\alpha_{8}}\,)$.

In some problems the situation appears where $\alpha_3=0$. For
this type of the problems, the solution given in Eq. (25) becomes
as

\begin{eqnarray}
\Psi(z)=z^{\alpha_{12}}e^{\alpha_{13}z}L^{\alpha_{10}-1}_{n}(\alpha_{11}z)\,,
\end{eqnarray}
and the energy spectrum is

\begin{eqnarray}
\alpha_{2}n-2\alpha_{5}n+(2n+1)(\,\sqrt{\alpha_{9}\,}-\alpha_{3}\sqrt{\alpha_{8}\,}\,)
&+&n(n-1)\alpha_{3}+\alpha_{7}\nonumber\\&+&2\alpha_{3}\alpha_{8}-2\sqrt{\alpha_{8}\alpha_{9}\,}+
\alpha_{5}=0\,.
\end{eqnarray}
when the limits become $lim_{\alpha_3 \rightarrow
0}P^{(\alpha_{10}-1\,,\,\frac{\alpha_{11}}{\alpha_3}-\alpha_{10}-1)}_{n}
(1-\alpha_{3}z)=L^{\alpha_{10}-1}_{n}(\alpha_{11}z)$ and
$lim_{\alpha_3 \rightarrow 0}
(1-\alpha_{3}z)^{-\alpha_{12}-\,\frac{\alpha_{13}}{\alpha_3}}=e^{\alpha_{13}z}$.

\section{Bound States}

\subsection{The Hypergeometric-Type Potential}

The Dirac equation has the exact pseudospin symmetry if
$\Sigma(r)=C=const.$\,, so Eq. (3) becomes under that condition

\begin{eqnarray}
\Big\{\frac{d^2}{dr^2}-\frac{\kappa(\kappa-1)}{r^2}-(m-E+C)M_{\Delta}(r)\Big\}G_{n\kappa}(r)=0\,,
\end{eqnarray}
where $\kappa=\tilde{\ell}+1$ for $\kappa>0$\,, and
$\kappa=-\tilde{\ell}$ for $\kappa<0$\,. The hypergeometric-type
potential is given by

\begin{eqnarray}
V(r)=D[1-\sigma \coth(\alpha
r)]^2=\Bigg(\frac{D_{1}+D_{2}e^{-2\alpha r}}{1-e^{-2\alpha
r}}\Bigg)^2\,,
\end{eqnarray}
where the real parameters $D$, $\sigma$, and $\alpha$ represent
the potential [22], and $D_{1}=\sqrt{D\,}(1-\sigma)$\,, and
$D_{2}=\sqrt{D\,}(1+\sigma)$.

Eq. (28) can not be solved analytically for any $\kappa$ values
because of $\kappa(\kappa-1)/r^2$ term, so we use the
approximation $1/r^2 \simeq 4\alpha^2e^{-2\alpha r}/(1-e^{-2\alpha
r})^2$ [26] to solve the equation for any spin-orbit quantum
number $\kappa$\,.

By using this approximation to centrifugal-like term, setting
$\Delta(r)$ to the potential given in Eq. (29), and inserting into
Eq. (28), we obtain

\begin{eqnarray}
\Big\{\frac{d^2}{dr^2}-4\alpha^2\kappa(\kappa-1)\frac{e^{-2\alpha
r}}{(1-e^{-2\alpha r})^2}+\mu\frac{(D_{1}+D_{2}e^{-2\alpha
r})^2}{(1-e^{-2\alpha r})^2}-\epsilon\Big\}G_{n\kappa}(r)=0\,,
\end{eqnarray}
where $\mu=m-E+C$\,, and $\epsilon=m(m+C)+E(C-E)$\,. At this
point, it is worthwhile to note that Eq. (29) becomes for
$\sigma=1 (D_{1}=0)$

\begin{eqnarray}
V(r)=D^2_{2}\Bigg(\frac{e^{-2\alpha r}}{1-e^{-2\alpha
r}}\Bigg)^2\,,
\end{eqnarray}

This form of the potential corresponds to the Manning-Rosen
potential for $A=0$ [27] if we set $\sqrt{\frac{1}{\kappa
b^2}\,\alpha(\alpha-1)\,} \rightarrow D_{2}$ (here, $\kappa$ and
$\alpha$ are the parameters in Ref. [27]), and $\frac{1}{b}
\rightarrow 2\alpha$\,. This means that we could also obtain the
energy eigenvalue equation of the Dirac equation for the
Manning-Rosen potential in the case of the exact spin symmetry, if
we set the parameter $D_{1}=0$ in the equations.

By using the new variable $z=e^{-2\alpha r} (0 < z < 1)$\,, we
obtain from Eq. (30)

\begin{eqnarray}
\frac{d^2G_{n\kappa}(z)}{dz^2}&+&\frac{1-z}{z(1-z)}\frac{dG_{n\kappa}(z)}{dz}+\frac{1}{[z(1-z)]^2}
\Big\{\beta^2(\mu
D^2_{1}-\epsilon)\nonumber\\&+&2\beta^2[D_{1}D_{2}\mu-\epsilon-2\alpha^2\kappa(\kappa-1)]
z+\beta^2(\mu D^2_{2}-\epsilon)z^2\Big\}G_{n\kappa}(z)=0\,.
\end{eqnarray}

By comparing Eq. (32) with Eq. (13), we get the parameter set

\begin{eqnarray}
\begin{array}{llll}
\alpha_{1}=1\,, & \alpha_{2}=1\,, & \xi_{1}=-\beta^2(\mu D^2_{2}-\epsilon), & \\
\alpha_{3}=1\,, & \alpha_{4}=0\,, & \xi_{2}=2\beta^2[D_{1}D_{2}\mu-\epsilon-2\alpha^2\kappa(\kappa-1)], & \\
\alpha_{5}=-\,\frac{1}{2}, & \alpha_{6}=\xi_{1}+\frac{1}{4}\,, &
\xi_{3}=-\beta^2(\mu D^2_{1}-\epsilon), & \\
\alpha_{7}=-\xi_{2}, & \alpha_{8}=\xi_{3}\,, & \alpha_{9}=\xi_{1}-\xi_{2}+\xi_{3}+\frac{1}{4}, &  \\
\alpha_{10}=1+2\sqrt{\xi_3}\,, &
\alpha_{11}=2+2(\,\sqrt{\xi_1-\xi_2+\xi_3+\frac{1}{4}\,}+\sqrt{\xi_3}\,),
& & \\ \alpha_{12}=\sqrt{\xi_3}\,, &
\alpha_{13}=-\frac{1}{2}-(\,\sqrt{\xi_1-\xi_2+\xi_3+\frac{1}{4}\,}+\sqrt{\xi_3}\,).
& &
\end{array}
\end{eqnarray}

The energy eigenvalue equation becomes

\begin{eqnarray}
\Bigg(\sqrt{\beta^2[4\alpha^2\kappa(\kappa-1)-\mu(D_{1}+D_{2})^2]+\frac{1}{4}\,}+\beta\sqrt{\epsilon-\mu
D^2_{1}\,}\Bigg) \Big(2n+1+2\beta\sqrt{\epsilon-\mu
D^2_{1}\,}\,\Big)\nonumber\\+\beta^2\Big[4\alpha^2\kappa(\kappa-1)-2(\epsilon
+\mu D_{1}D_{2})\Big]=-n(n+1)-1/2\,.
\end{eqnarray}

In this case, we use only the negative energy eigenvalues, because
negative energy states exist in the pseudospin symmetry [28].

Now, let us give the corresponding Dirac spinors from Eq. (25)

\begin{eqnarray}
G_{n\kappa}(z)&=&z^{\beta\sqrt{\epsilon-\mu D^2_{1}\,}}\nonumber\\
&&\times(1-z)^{\frac{1}{2}\,+\sqrt{-\mu\beta^2(D_{1}+D_{2})^2
+\kappa(\kappa-1)+\frac{1}{4}\,}} P_{n}^{(2\beta\sqrt{\epsilon\mu
D^2_{1}\,}\,,\,2\sqrt{-\mu\beta^2(D_{1}+D_{2})^2
+\kappa(\kappa-1)+\frac{1}{4}\,}\,)}(1-2z)\,.\nonumber\\
\end{eqnarray}

Finally, we briefly give the energy eigenvalue equation for the
special case $\sigma=1$\,, which gives the energy spectra of the
Manning-Rosen potential with $A=0$

\begin{eqnarray}
\Bigg(\sqrt{4\beta^2[\alpha^2\kappa(\kappa-1)-D\mu]+\frac{1}{4}\,}+\beta\sqrt{\epsilon\,}\Bigg)
\Big(2n+1+2\beta\sqrt{\epsilon\,}\,\Big)\nonumber\\+4\beta^2\Big[\alpha^2\kappa(\kappa-1)-
\frac{\epsilon}{2}\Big]+\,\frac{1}{4}\,((2n+1)^2+1)=0\,.
\end{eqnarray}

Under the exact spin symmetry, i.e. $\Delta(r)=C=const.$\,, Eq.
(4) becomes

\begin{eqnarray}
\Big\{\frac{d^2}{dr^2}-\frac{\kappa(\kappa+1)}{r^2}-(m+E-C)M_{\Sigma}(r)\Big\}F_{n\kappa}(r)=0\,,
\end{eqnarray}
where $\kappa=-(\ell+1)$ for $\kappa>0$\, and $\kappa=\ell$ for
$\kappa<0$\,. By setting $\Sigma(r)$ to the potential given in Eq.
(30), using the above approximation to the $\kappa(\kappa+1)/r^2$
term, and using the new variable $z=e^{-2\alpha r} (0 < z < 1)$,
we obtain

\begin{eqnarray}
\frac{d^2F_{n\kappa}(z)}{dz^2}&+&\frac{1-z}{z(1-z)}\frac{dF_{n\kappa}(z)}{dz}+\frac{1}{[z(1-z)]^2}
\Big\{\beta^2(\epsilon'-\mu'D^2_{1})\nonumber\\&-&\beta^2[2\epsilon'+2D_{1}D_{2}\mu'+4\alpha^2\kappa(\kappa+1)]
z+\beta^2(\epsilon'-\mu'D^2_{2})z^2\Big\}F_{n\kappa}(r)=0\,.
\end{eqnarray}
where $\mu'=m+E-C$, $\epsilon'=m(C-m)+E(E-C)$, and
$\beta^2=1/4\alpha^2$. By comparing Eq. (39) with Eq. (13), we
obtain the parameter set given in Eq. (34) where
$\xi_{1}=-\beta^2(\epsilon'-\mu'D^2_{2})$,
$\xi_{2}=-2\beta^2[\epsilon'+D_{1}D_{2}\mu'+2\alpha^2\kappa(\kappa+1)]$,
and $\xi_{3}=-\beta^2(\epsilon'-\mu'D^2_{1})$. The energy
eigenvalue equation of the hypergeometric potential for the exact
spin symmetry is written from Eq. (21) as

\begin{eqnarray}
\Bigg(\sqrt{\beta^2\mu'(D_{1}+D_{2})^2+\kappa(\kappa+1)+\frac{1}{4}\,}+\beta\sqrt{\mu'D^2_{1}-\epsilon'\,}\Bigg)
\Big(2n+1+2\beta\sqrt{\mu'D^2_{1}-\epsilon'\,}\,\Big)\nonumber\\+\beta^2\Big[4\alpha^2\kappa(\kappa+1)+2(\epsilon'
+\mu'D_{1}D_{2})\Big]=-n(n+1)-1/2\,,
\end{eqnarray}

The last equation can give negative, and positive eigenvalues, but
we choose only positive energy eigenvalues, because in the case of
the exact spin symmetry appears only the positive energy
eigenstates [28]. The corresponding Dirac spinors are obtained
from Eq. (25), and given

\begin{eqnarray}
F_{n\kappa}(z)&=&z^{\beta\sqrt{\mu'D^2_{1}-\,\epsilon'\,}}\nonumber\\
&&\times(1-z)^{\frac{1}{2}\,+\sqrt{\mu'\beta^2(D_{1}+D_{2})^2
+\kappa(\kappa+1)+\frac{1}{4}\,}}
P_{n}^{(2\beta\sqrt{\mu'D^2_{1}-\,\epsilon'\,}\,,\,2\sqrt{\mu'\beta^2(D_{1}+D_{2})^2
+\kappa(\kappa+1)+\frac{1}{4}\,}\,)}(1-2z)\,.\nonumber\\
\end{eqnarray}

Now, we briefly give the energy eigenvalue equation for the
Manning-Rosen potential with $A=0$ ($\sigma=1$)

\begin{eqnarray}
&&\Bigg(\sqrt{4\beta^2D(m+E-C)+\kappa(\kappa+1)+\frac{1}{4}\,}+\beta\sqrt{m(m-C)+E(C-E)\,}\Bigg)\nonumber\\
&\times&\Big(2n+1+2\beta\sqrt{m(m-C)+E(C-E)\,}\,\Big)\nonumber\\&+&\beta^2\Big[4\alpha^2\kappa(\kappa+1)+
2(m(C-m)+E(E-C))\Big]+\,\frac{1}{4}\,((2n+1)^2+1)=0\,.
\end{eqnarray}

\subsection{The Generalized Morse Potential}

Assuming that the potential $\Delta(r)=V_v(r)+V_s(r)$ is the
generalized Morse potential given by

\begin{eqnarray}
V(r)=V_{1}e^{-2\alpha r}-V_2 e^{-\alpha r}\,,
\end{eqnarray}
and substituting Eq. (42) into Eq. (3), and taking
$\Sigma(r)=\Sigma=const.$, we have the following equation in the
exact pseudospin symmetry for $\kappa=0$ ($z=e^{-\alpha r}$)

\begin{eqnarray}
\Big\{\frac{d^2}{dz^2}+\frac{1}{s}\frac{d}{dz}&+&\frac{1}{z^2}\Big[4\beta^2\Big(\mu^2-E^2+E(\mu+\Sigma)\Big)\nonumber\\
&+&4\beta^2V_2(\mu-E+\Sigma)z+4\beta^2V_1(E-\mu-\Sigma)z^2\Big]\Big\}G_{n\kappa}(z)=0\,.
\end{eqnarray}

Comparing Eq. (43) with Eq. (13), we obtain the following
parameter set

\begin{eqnarray}
\begin{array}{llll}
\alpha_1=1\,, & \alpha_2=0, & \xi_1=4\beta^2V_1(\mu-E+\Sigma), & \\
\alpha_3=0\,, & \alpha_4=0, & \xi_2=4\beta^2V_2(\mu-E+\Sigma), & \\
\alpha_5=0, & \alpha_6=\xi_1, &
\xi_3=4\beta^2(E^2-\mu^2-E(\mu+\Sigma)), & \\
\alpha_7=-\xi_2, & \alpha_8=\xi_3\,, & \alpha_9=\xi_1 &  \\
\alpha_{10}=1+2\sqrt{\xi_3}\,, & \alpha_{11}=2\sqrt{\xi_1\,}, & \\
\alpha_{12}=\sqrt{\xi_3}\,, & \alpha_{13}=-\sqrt{\xi_1\,}.
\end{array}
\end{eqnarray}

From Eq. (27), we obtain the energy eigenvalue equation for
$\kappa=0$

\begin{eqnarray}
E^2-E(\mu+\Sigma)-\mu^2=\frac{1}{16\beta^2}\Bigg(2n+1-\frac{2\beta
V_2}{\sqrt{V_1\,}}\sqrt{\mu-E+\Sigma\,}\Bigg)^2\,,
\end{eqnarray}
where $\beta^2=1/4\alpha^2$. We should choose the negative energy
solution in Eq. (46) because the negative energy states exist only
in the exact pseudospin limit. The corresponding lower spinor
component can be obtained from Eq. (26)

\begin{eqnarray}
G_{n\kappa}(z)&=&z^{2\beta\sqrt{E^2-\mu^2-E(\mu+\Sigma)\,}}\,e^{-2\beta\sqrt{V_1(\mu-E+\Sigma)\,}\,z}
\nonumber\\&\times&L^{4\beta\sqrt{E^2-\mu^2-E(\mu+\Sigma)\,}}_{n}\,(4\beta\sqrt{V_1(\mu-E+\Sigma)\,}\,z)\,,
\end{eqnarray}

In the case of exact spin symmetry the potential
$\Delta(r)=V_{v}(r)-V_{s}(r)$ is a constant, let say
$\Delta(r)=\Delta=const.$\,. We set the potential $\Sigma(r)$ as
the Morse potential in Eq. (42). Substituting the potential into
Eq. (4), and using the same variable $z=e^{-\alpha r}$, we obtain

\begin{eqnarray}
\Big\{\frac{d^2}{dz^2}+\frac{1}{z}\frac{d}{dz}&+&\frac{1}{z^2}\Big[4\beta^2\Big(E^2-\mu^2+
\Delta(\mu-E)\Big)\nonumber\\
&+&4\beta^2V_2(\mu+E-\Delta)z+4\beta^2V_1(-E-\mu+\Delta)z^2\Big]\Big\}F_{n\kappa}(z)=0\,,
\end{eqnarray}

Comparing Eq. (47) with Eq. (13), we obtain the parameter set
given in Eq. (44) where $\xi_1=4\beta^2V_1(E+\mu-\Delta),
\xi_2=4\beta^2V_2(E+\mu-\Delta)$ and
$\xi_3=4\beta^2(\mu^2-E^2+\Delta(E-\mu))$. From Eq. (27), we
obtain the energy eigenvalue equation in the case of exact spin
symmetry for $\kappa=0$

\begin{eqnarray}
\mu^2-E^2+\Delta(E-\mu)=\frac{1}{16\beta^2}\Bigg(2n+1-\frac{2\beta
V_2}{\sqrt{V_1\,}}\sqrt{E+\mu-\Delta\,}\Bigg)^2\,.
\end{eqnarray}
where $\beta^2=1/4\alpha^2$. We should choose the positive energy
solution in Eq. (49) because the positive energy states exist only
in the exact spin limit. The corresponding Dirac spinor can be
written as

\begin{eqnarray}
F_{n\kappa}(z)&=&z^{2\beta\sqrt{\mu^2-E^2+\Delta(E-\mu)\,}}\,e^{-2\beta\sqrt{V_1(\mu+E-\Delta)\,}\,z}
\nonumber\\&\times&L^{4\beta\sqrt{\mu^2-E^2+\Delta(E-\mu)\,}}_{n}\,(4\beta\sqrt{V_1(\mu+E-\Delta)\,}\,z)\,.
\end{eqnarray}

\subsection{The Pöschl-Teller Potential}

By taking the potential $\Delta(r)=V_v(r)+V_s(r)$ is the
P\"{o}schl-Teller potential [23] given by

\begin{eqnarray}
V(r)=-4V_0\frac{e^{-2\alpha r}}{(1+e^{-2\alpha r})^2}\,,
\end{eqnarray}
and substituting Eq. (50) into Eq. (3), taking into account
$\Sigma(r)=\Sigma=const.$, we have the following equation in the
exact pseudospin symmetry for $\kappa=0$ ($z=-e^{-2\alpha r}$)

\begin{eqnarray}
\frac{d^2G_{n\kappa}(z)}{dz^2}&+&\frac{1-z}{z(1-z)}\frac{dG_{n\kappa}(z)}
{dz}\nonumber\\&+&\frac{1}{[z(1-z)]^2}\Big\{
\beta^2(\mu-E+\Sigma)\Big[\mu+E-[2\mu+2E+4V_0]z\nonumber\\&+&(\mu+E)z^2\Big]\Big\}G_{n\kappa}(z)=0\,.
\end{eqnarray}

Following the same procedure, we obtain the parameter set

\begin{eqnarray}
\begin{array}{ll}
\alpha_1=1\,, & \xi_1=\beta^2(\mu+E)(-\mu-\Sigma+E),\\
\alpha_2=1\,, &
\xi_2=2\beta^2(-\mu-\Sigma+E)[\mu+E+2V_0]\,, \\
\alpha_3=1\,, &
\xi_3=\beta^2(\mu+E)(-\mu-\Sigma+E)\,, \\ \alpha_4=0\,, & \alpha_5=-\frac{1}{2}, \\
\alpha_6=\xi_1+\frac{1}{4}\,, & \alpha_7=-\xi_2, \\
\alpha_8=\xi_3\,, & \alpha_9=\xi_1-\xi_2+\xi_3+\frac{1}{4}, \\
\alpha_{10}=1+2\sqrt{\xi_3}\,, & \alpha_{11}=2+2(\,\sqrt{\xi_1-\xi_2+\xi_3+\frac{1}{4}\,}+\sqrt{\xi_3\,}\,), \\
\alpha_{12}=\sqrt{\xi_3}\,, &
\alpha_{13}=-\frac{1}{2}-(\,\sqrt{\xi_1-\xi_2+\xi_3+\frac{1}{4}\,}+\sqrt{\xi_3\,}\,)\,.
\end{array}
\end{eqnarray}
and the energy eigenvalue equation of P\"{o}schl-Teller potential
under the exact pseudospin symmetry for $\kappa=0$ from Eq. (21)

\begin{eqnarray}
E^2-\mu^2-\Sigma(\mu+E)=\frac{1}{4}\bigg((2n+1)\alpha+\sqrt{4V_0(\mu-E+\Sigma)+\alpha^2\,}\bigg)^2\,.
\end{eqnarray}

The last energy eigenvalue equation has a quadratic form in terms
of energy $E$. We take the negative energy vales in the exact
pseudospin limit. The corresponding Dirac spinor can be written in
terms of Jacobi polynomials, i.e., $P^{(\alpha\,,\,\beta)}_n (x)$,

\begin{eqnarray}
G_{n\kappa}(z)&=&z^{\beta\sqrt{(\mu+E)(-\mu-\Sigma+E)\,}}\,(1-z)^{\frac{1}{2}\Big[\,1+\sqrt{1+16V_0
\beta^2(\mu+\Sigma-E)\,}\,\Big]}\nonumber\\&\times&P^{(2\beta\sqrt{(\mu+E)(-\mu-\Sigma+E)\,}\,,\,\sqrt{1+16V_0
\beta^2(\mu+\Sigma-E)\,}\,)}_{n}\,(1-2z)\,,
\end{eqnarray}

In the case of exact spin symmetry, we set the potential
$\Sigma(r)$ as P\"{o}schl-Teller potential given in Eq. (50),
$\Delta(r)=\Delta=const.$, and by using the coordinate
transformation $z=-e^{-2\alpha r}$, we obtain from Eq. (4)

\begin{eqnarray}
\frac{d^2F_{n\kappa}(z)}{dz^2}&+&\frac{1-z}{z(1-z)}
\frac{dF_{n\kappa}(z)}{dz}\nonumber\\&+&\frac{\Delta-\mu-E}{[z(1-z)]^2}\Big\{
\beta^2(\mu-E)-2\beta^2(\mu-E-2V_0)z\nonumber\\&+&\beta^2(\mu-E)z^2\Big\}F_{n\kappa}(z)=0\,,
\end{eqnarray}
which gives the parameter set given in Eq. (52) where
$\xi_1=\beta^2(E-\mu)(\Delta-\mu-E)$,
$\xi_2=2\beta^2(\Delta-\mu-E)(E-\mu+2V_0)$ and
$\xi_3=\beta^2(E-\mu)(\Delta-\mu-E)$. The energy eigenvalue
equation of the P\"{o}schl-Teller potential under the exact spin
symmetry for $\kappa=0$ from Eq. (21) is obtained

\begin{eqnarray}
E^2+\mu^2-\Delta(\mu-E)=\frac{1}{4}\bigg((2n+1)\alpha+\sqrt{4V_0(\mu+E-\Delta)+\alpha^2\,}\bigg)^2\,.
\end{eqnarray}

Finally, we obtain the corresponding Dirac spinor from Eq. (25)

\begin{eqnarray}
F_{n\kappa}(z)&=&z^{\beta\sqrt{(E-\mu)(\Delta-\mu-E)\,}}\,(1-z)^{\frac{1}{2}\Big[\,1+\sqrt{1+16V_0
\beta^2(\mu+E-\Delta)\,}\,\Big]}\nonumber\\&\times&P^{(2\beta\sqrt{(E-\mu)(\Delta-\mu-E)\,}\,,\,\sqrt{1+16V_0
\beta^2(\mu+E-\Delta)\,}\,)}_{n}\,(1-2z)\,.
\end{eqnarray}

\section{Conclusion}

We have studied the energy eigenvalues and the corresponding
eigenfunctions of the Dirac equation with the hypergeometric
potential, the Morse potential, and the Pöschl-Teller potential in
the case of pseudospin, and spin symmetry. We have used the
parametric generalization of the NU method to obtain the results.
The energy eigenvalues of all potentials are real and the wave
functions are written in terms of the Laguerre (Jacobi)
polynomials. We have also investigated the special case $\sigma=1$
in the case of the hypergeometric potential, which corresponds to
the case of the Manning-Rosen potential with $A=0$\,. So we have
obtained the energy eigenvalue equation of the Manning-Rosen
potential in the pseudospin, and spin symmetry case, respectively.

\end{document}